\documentclass[a4paper,11pt]{article}
\usepackage{pos}

\usepackage{graphicx}
\usepackage{subcaption}
\usepackage{caption}

\title{Detection Prospects for AGNs with the Cherenkov Telescope Array}

\author*[a]{Luana Passos-Reis}
\author[a]{Elisabete M. de Gouveia Dal Pino}
\author[b]{Tarek Hassan}
\author[c]{Santiago Pita}
\author[d,e]{Jonathan Biteau}
\author[f]{Jean-Philippe Lenain}
\author[g]{Atreya Acharyya}
\author[h]{Alberto Domínguez}
\author[i]{Lucas Gréaux}
\author[j]{Luiz Augusto Stuani Pereira}
\author[j]{Edivaldo Moura Santos}
\author[]{Paolo Goldoni$^{k}$, for the CTAO Consortium}

\affiliation[a]{Instituto de Astronomia, Geof\'{i}sica e Ci\^{e}ncias Atmosf\'{e}ricas (IAG-USP), Universidade de S\~{a}o Paulo, \\
Rua do Mat\~{a}o 1226, CEP: 05508-090, S\~{a}o Paulo - SP, Brazil.}

\affiliation[b]{Centro de Investigaciones Energéticas, Medioambientales y Tecnológicas (CIEMAT), 40, Madrid, Spain.}

\affiliation[c]{Université Paris Cité, CNRS, Astroparticule et Cosmologie, F-75013 Paris, France.}

\affiliation[d]{Universit\'e Paris-Saclay, CNRS/IN2P3, IJCLab, Orsay, France.}

\affiliation[e]{Institut universitaire de France (IUF).}

\affiliation[f]{Laboratoire de Physique Nucléaire et de Hautes Energies (LPNHE), Sorbonne Université, CNRS/IN2P3, Paris, France.}

\affiliation[g]{CP3-Origins, University of Southern Denmark, Campusvej 55, 5230 Odense M, Denmark.}

\affiliation[h]{IPARCOS and Department of EMFTEL, Universidad Complutense de Madrid, E-28040 Madrid, Spain.}

\affiliation[i]{Fakultät für Physik \& Astronomie, Ruhr-Universität Bochum, D-44780 Bochum, Germany.}

\affiliation[j]{Instituto de Física (IFUSP), Universidade de S\~{a}o Paulo, Rua do Mat\~{a}o 1371, CEP: 05508-090, S\~{a}o Paulo - SP, Brazil.}

\affiliation[k]{Université Paris Cité, CNRS, CEA, Astroparticule et Cosmologie, F-75013 Paris, France.}

\emailAdd{$^{*}$luana.passos.reis@usp.br}

\abstract{The Cherenkov Telescope Array Observatory (CTAO) will enable detailed studies of Active Galactic Nuclei (AGN) in the very-high-energy (VHE) regime, as the next-generation ground-based gamma-ray observatory, designed to enhance sensitivity and energy coverage (20 GeV -- 300 TeV) over current Imaging Atmospheric Cherenkov Telescopes (IACTs). In the context of the CTAO Science Collaboration, within the AGN Population working group, we developed a variability-based strategy to improve predictions of AGNs detectable by CTAO, using Fermi-LAT data and normalized excess variance (NXS) as a tracer of flux variability. By extrapolating from 30-day to 3-day timescales, we expanded the sample of sources with short-timescale variability estimates from 87 to 407. This approach allows us to identify flaring and distant AGNs that are promising CTAO targets. The results are being used to support the CTAO extragalactic science program and will be included in an upcoming Consortium publication for the AGN Population collaboration.}

\ConferenceLogo{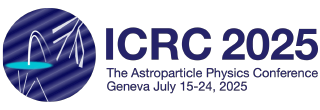}

\FullConference{39th International Cosmic Ray Conference (ICRC2025)\\
 15–24 July 2025\\
Geneva, Switzerland\\}

\begin{document}
\maketitle

\section{Introduction}

To estimate the extragalactic sky observable by CTAO, we use the Third Data Release of the Fourth Fermi-LAT Catalog of Active Galactic Nuclei (4LAC-DR3) sources \cite{Ajello_2022}, which provides detailed spectral and variability information for 3814 AGNs. The catalog includes major blazar subclasses, BL Lacs and FSRQs, as well as other radio-loud AGNs, allowing for a classification-based population analysis.

Blazars, a subclass of active galactic nuclei (AGN) with relativistic jets aligned towards our line of sight, dominate the extragalactic gamma-ray sky. Their broadband non-thermal emission is highly variable, especially at high energies. While current Imaging Atmospheric Cherenkov Telescopes (IACTs) have advanced our understanding of these sources, their limited field of view and duty cycle restrict the ability to build a complete picture of the TeV extragalactic sky yet.

The upcoming Cherenkov Telescope Array Observatory (CTAO), with its enhanced sensitivity and broad energy coverage (20 GeV – 300 TeV),
will offer a major step forward in AGN studies. To anticipate which sources CTAO might detect, we rely on the well-established GeV sky from Fermi-LAT and extrapolate it to the very-high energy regime. This effort, as part of the CTAO AGN Population Task Force, aims to refine population models by incorporating both spectral characteristics and variability trends observed in current data.

In this work, we focus on the variability-based source selection scheme and the role of short-term variability in improving CTAO’s work within the AGN Population task force. A full population and spectral analysis will be presented in our forthcoming collaboration paper (see also \cite{Hassan_2017}).


\section{Extragalactic Source Population Studies}

The primary goal of the presented work is detailed in an upcoming CTAO Consortium publication, where a realistic estimate of the population of extragalactic sources accessible to the CTAO is provided, based on 5-hour and 20-hour observational strategies. These forecasts are stratified by AGN subclass and redshift, and include sky maps highlighting currently known TeV AGNs. To generate these predictions, the study extrapolates the well-characterized GeV sky observed by Fermi-LAT into the TeV regime, applying spectral assumptions consistent with current Imaging Atmospheric Cherenkov Telescopes (IACT) observations.

While current IACTs have significantly improved our understanding of AGN variability and broadband emission, their limited field of view (typically $\sim$4 degrees in diameter) has constrained efforts to build an unbiased and complete view of the TeV extragalactic sky. Our forecasts rely on quiescent-state fluxes derived from Fermi-LAT data and therefore represent conservative lower limits on CTAO’s detection capabilities.


However, the highly variable nature of blazars, particularly on timescales relevant for CTAO, suggests that many sources may become significantly brighter during flaring episodes. In this work, we focus on characterizing this variability using the normalized excess variance (NXS) to identify promising flaring AGN candidates. This variability-based subselection of sources serves as a key input to incorporate observational constraints and refine detection forecasts. This inclusion reveals a substantial increase in the expected number of detectable sources, offering a more dynamic and realistic view of the extragalactic gamma-ray sky that CTAO will explore.

\section{Processing and Selection of Light Curves}

The available light curve sample from Fermi-LAT \cite{FermiLCR} (see also \cite{Ajello_2022}), which includes 1429 AGNs, 
provides a robust dataset for temporal variability analysis across multiple timescales. These sources form the foundation for evaluating the impact of variability on CTAO detectability using flux measurements binned over 3-day, 7-day (weekly), and 30-day (monthly) intervals. We retrieved these light curves with a minimum detection significance of $TS = 1$, where any bins with $TS < 1$ represent upper limits instead of flux measurements.

To estimate variability parameters for each AGN, we applied a quality filter to individual light curve bins. Data points that did not satisfy the following criteria were flagged as \textit{unconstrained points} and subsequently tagged for replacement of their flux values in the variability analysis, as detailed later in this section. These criteria are:

\begin{itemize}
\item Test Statistic (TS) value below a threshold: TS $<$ 2;
\item Zero uncertainty on the flux value: flux\_error = 0;
\item Non-convergent likelihood fits: fit\_convergence $\neq$ 0;
\item Insufficient exposure: exposure $<$ $1 \times 10^7$ cm$^{2}$ s;
\item Invalid flux measurements: flux = NaN;
\item Missing upper limits where flux = NaN: flux\_UL = NaN;
\item Non-physical upper limits: flux\_UL $<$ 0 for bins with NaN flux.
\end{itemize}

\begin{figure}[ht]
  \centering
  \includegraphics[width=0.75\textwidth]{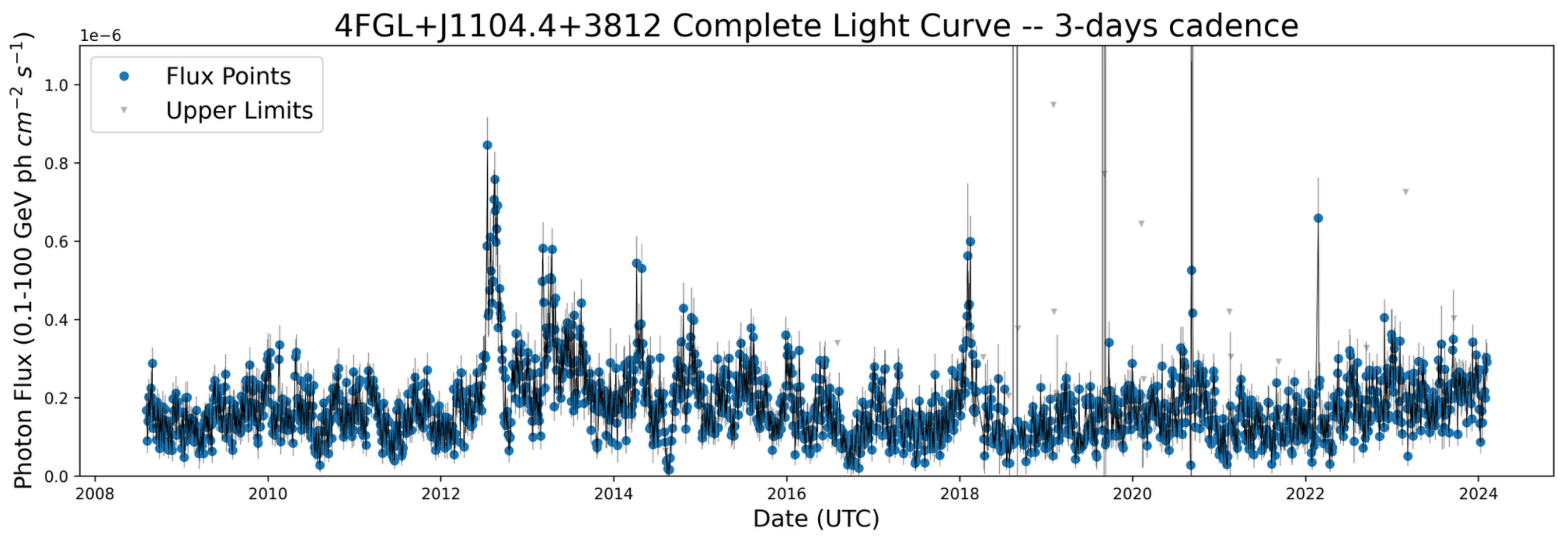}
  \caption{Unfiltered 3-day light curve of Mkn 421. Outliers and unconstrained flux points appear as extreme deviations, including vertical lines that extend beyond the graph boundaries.}
  \label{fig:LC_complete_Mkn421}
\end{figure}

\begin{figure}[ht]
  \centering
  \includegraphics[width=0.75\textwidth]{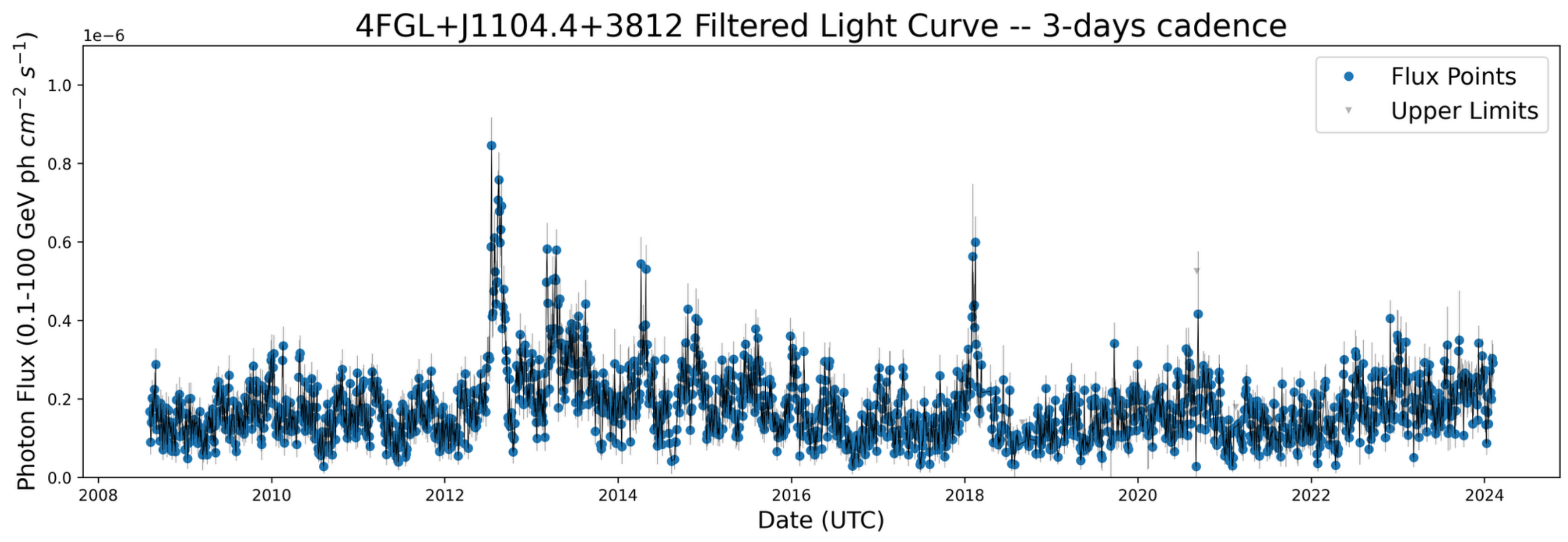}
  \caption{Filtered 3-day light curve of Mkn 421, after removing outliers and unconstrained points.}
  \label{fig:LC_filtered_Mkn421}
\end{figure}

As an illustrative example, Figures \ref{fig:LC_complete_Mkn421} and \ref{fig:LC_filtered_Mkn421} display the 3-day light curve of the well-known AGN Mkn 421, before and after the removal of outliers and unconstrained flux points. In the unfiltered light curve, a few flux values appear significantly higher than the bulk of the data, often by orders of magnitude. These anomalies are likely caused by poor exposure, low detection significance, or large flux uncertainties. Since our goal is to characterize the long-term variability of bright and consistently detected sources, such outliers can introduce significant bias in the analysis and must be carefully filtered.

Following these initial filtering steps, we further refine the light curve dataset by excluding sources that lack a sufficient number of valid flux measurements. To quantify the relative presence of usable data, we define a \textbf{quality ratio} as:

\begin{equation}
    \text{quality ratio} = \frac{\text{number of upper limits + unconstrained points}}{\text{number of actual flux measurements}}.
\end{equation}

Sources with a quality ratio less than 1 are retained in the analysis, a selection criterion that ensures at least half of the light curve bins contain meaningful flux detections. We also identify the total number of invalid bins after filtering.

For the selected sources, we replaced the flux and flux error values of the unconstrained bins as follows: the flux was set to 0, and the corresponding flux error is then filled using the value from the previous valid flux bin (forward-filling). However, if a valid bin is not present at the beginning of the light curve, we fill the initial missing flux error with the median flux value derived from all bins with a Test Statistic (TS) between 1 and 2. This procedure ensures a consistent light curve structure and minimizes the impact of unreliable measurements and any artificial bias introduced by unconstrained flux points.




We have validated this approach by performing a cross-check analysis (Passos Reis et al., in prep.) with the History Flux parameter and the Fractional Variability (available in \cite{Ajello_2022}), supporting its use for handling both unconstrained bins and missing values in the Normalized Excess Variance (NXS) estimation below.

\section{Selecting Blazar Targets based on Variability}\label{variability} 

To identify promising blazar targets for the Cherenkov Telescope Array Observatory (CTAO), we constructed a filtered and variability-selected subsample of the 4LAC catalog based on Fermi-LAT light curves. This approach focuses on sources with sufficient brightness to derive reliable light curve statistics and Spectral Energy Distributions (SEDs). For each source, we compute key variability indicators, including the Fractional Variability ($F_{\rm var}$) and the Normalized Excess Variance ($\sigma_{\rm NXS}^2$), which capture the amplitude of flux variation over time.

These parameters serve as proxies for spectral activity and temporal flux variations, enabling extrapolation of average and flaring behavior into the VHE gamma-ray band.

\subsection*{Introducing the Normalized Excess Variance Parameter (NXS): $\sigma_{\rm NXS}^{2} \pm \Delta \sigma_{\rm NXS}^{2}$}

The normalized excess variance captures the amplitude of intrinsic variability, which is used to guide our flux extrapolations. In particular, we incorporate NXS into our variable sky modeling for CTAO by scaling the average flux with a variability term that preserves the observed amplitude on a given timescale. This allows us to simulate not just average but flare-enhanced VHE fluxes from Fermi-detected AGNs.

This behavior is statistically quantified via the normalized excess variance $\sigma_{\rm NXS}^{2}$ \cite{Vaughan_2003}:

\begin{equation}\label{eq:nxs}
    \sigma_{\rm NXS}^{2} = \frac{1}{F_{av}\ ^{2}} \left [\frac{1}{N - 1} \sum_{i=1}^{N} \left ( F_{i} - F_{av} \right )^{2} - \frac{1}{N} \sum_{i=1}^{N} \sigma_{\rm err, i}^{2} \right ],
\end{equation}

with uncertainty

\begin{equation}
    err(\sigma_{\rm NXS}^{2}) = \Delta \sigma_{\rm NXS}^{2} = \sqrt{\left ( \sqrt{\frac{2}{N}} \cdot \frac{\overline{\sigma_{\rm err}^{2}}}{F_{av}\ ^{2}} \right )^{2} + \left ( \sqrt{\frac{\overline{\sigma_{\rm err}^{2}}}{N}} \cdot \frac{2 F_{\rm var}}{F_{av}} \right )^{2}},
\end{equation}

where $N$ is the number of valid flux measurements, $F_{i}$ the flux in bin $i$, $F_{\rm av}$ the average flux, and $\overline{\sigma_{\rm err}^{2}}$ the mean squared flux error. The term $F_{\rm var}$ represents the fractional variability amplitude and is defined as $F_{\rm var} = \max{ \left( 0, \sqrt{\sigma_{\rm NXS}^{2}} \right) }$, ensuring that $F_{\rm var}$ is set to zero when $\sigma_{\rm NXS}^{2}$ is negative due to statistical fluctuations.

\begin{figure}[ht]
  \centering
  \begin{subfigure}[t]{\textwidth}
    \centering
    \includegraphics[width=\textwidth]{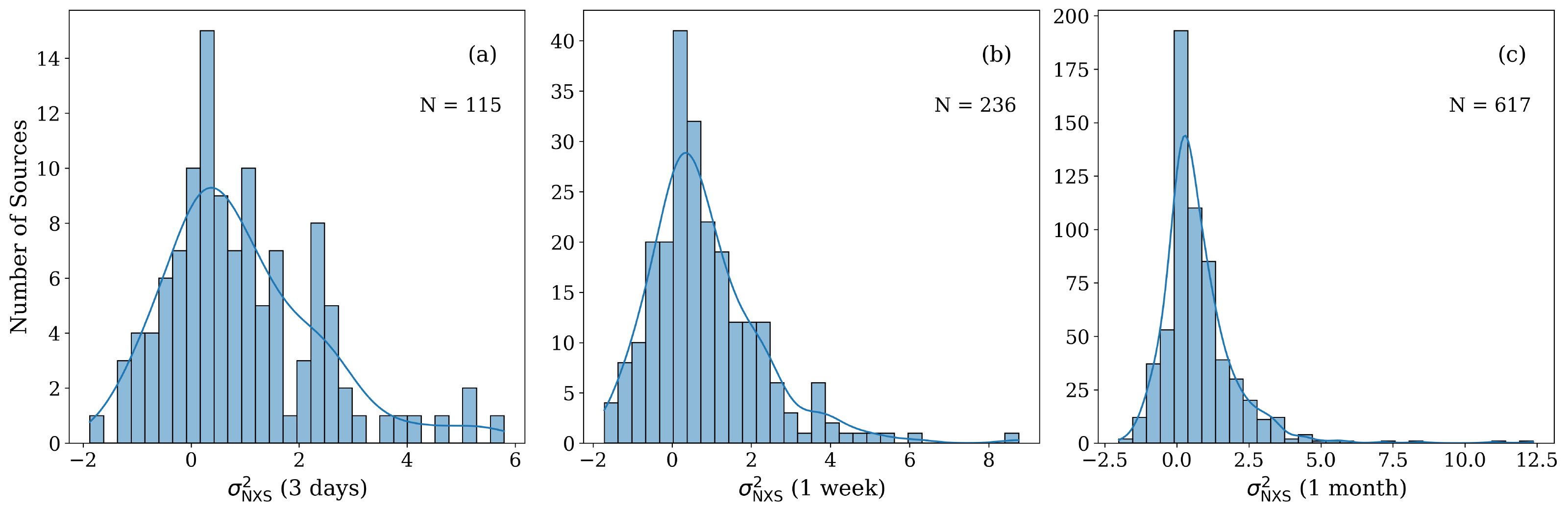}
    \caption{Distribution of $\sigma_{\rm NXS}^{2}$ for sources passing flux quality and 50\% data completeness filters. We retain 115 (3-day), 236 (weekly), and 617 (monthly) sources from the initial 1429 dataset.}
  \end{subfigure}
  \begin{subfigure}[t]{\textwidth}
    \centering
    \includegraphics[width=0.8\textwidth]{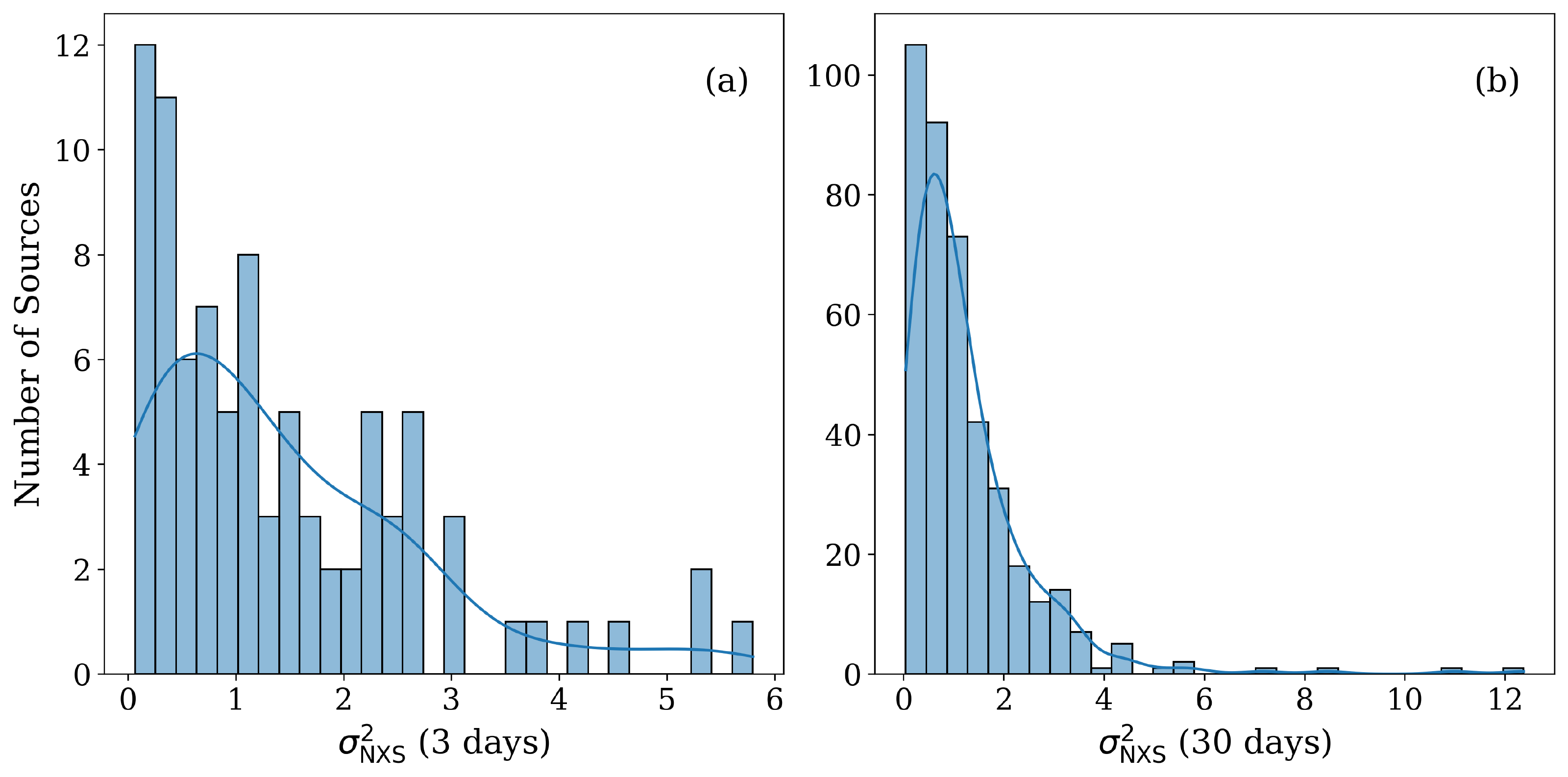}
    \caption{Subset of most variable sources after applying a 3$\sigma$ filter on $\sigma_{\rm NXS}^{2}$, with 87 (3-day) and 407 (monthly) light curves selected.}
  \end{subfigure}
  \caption{Normalized excess variance $\sigma_{\rm NXS}^{2}$ (NXS) distributions across different cadences, after applying completeness and variability filters.}
  \label{fig:gaps_distributions}
\end{figure}

Figure \ref{fig:gaps_distributions} shows the distributions of $\sigma_{\rm NXS}^{2}$ across the 3-day, weekly, and monthly cadences, after applying a 50\% completeness threshold, requiring that at least half of the bins contain measured fluxes rather than upper limits or missing values. This filtering step reduces the initial set of 1429 AGNs to a more reliable subset of 617 monthly, 236 weekly, and 115 3-day light curves, providing a cleaner basis for variability analysis.

In the second panel, we further refine this sample by selecting only those sources with significant variability, defined by $\sigma_{\rm NXS}^{2} > 3 \Delta \sigma_{\rm NXS}^{2}$. This leads to a final subset of 407 monthly and 87 sources in a 3-day binning, used in further correlation and extrapolation analyses. These represent the most variable and well-sampled AGNs, forming a promising sub-catalog of the Fourth LAT AGN Catalog (4LAC), optimized for variability analysis and extrapolation to the VHE regime relevant to CTAO.

Note that these numbers depend on the filtering parameters applied (e.g., TS thresholds, minimum exposure, or treatment of upper limits) and may vary accordingly.

The final dataset allows us to characterize the variability landscape across cadences and assess its evolution from long (monthly) to short (3-day) timescales. These sources are likely to exhibit detectable flux enhancements in the CTAO's observational window.
\section{Determination of the proportionality coefficient between $\sigma_{\rm NXS, 3d}^{2} (\text{NXS 3-days})$ and $\sigma_{\rm NXS, 30d}^{2} (\text{NXS 30 days})$}

To empirically connect short and long timescale variability, we selected the most variable AGNs on monthly cadence, defined by $\sigma_{\rm NXS, 30d}^{2}$ exceeding the 3$\sigma$ threshold. This resulted in a subsample of 407 sources. Among these, 87 had reliable light curves on both monthly and 3-day binnings after applying strict quality filters (50\% data gaps, removal of unconstrained points, and 3$\sigma$ significance cuts).

\begin{figure}[ht]
  \centering
  \includegraphics[width=0.75\textwidth]{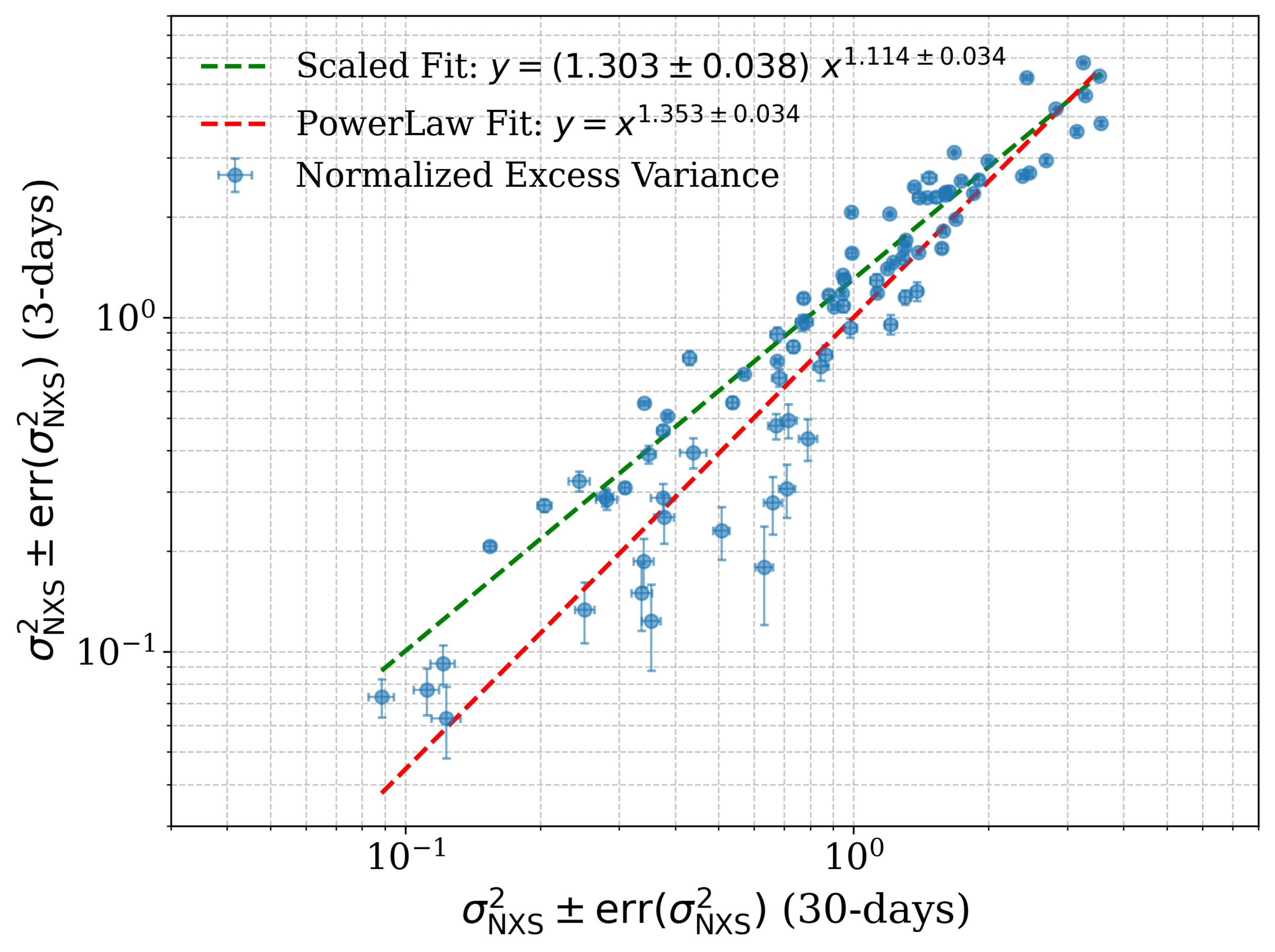}
  \caption{Correlation between $\sigma_{\rm NXS}^{2}$ on 3-day and 1-month timescales for a sample of 87 AGNs with high-significance variability. The fits enable extrapolation of short-timescale variability from monthly data, providing a practical approach to extend temporal coverage for population studies.}
  \label{fig:fit_NXS}
\end{figure}

As shown in Figure~\ref{fig:fit_NXS}, a clear near-linear correlation emerges between the $\sigma_{\rm NXS}^{2}$ values at the two timescales. A scaled fit yields $\sigma_{\rm NXS,3d}^{2} \approx (1.303 \pm 0.038) \cdot ( \sigma_{\rm NXS,30d}^{2} )^{1.114 \pm 0.034}$, while the power-law fit (an unscaled version) gives $\sigma_{\rm NXS,3d}^{2} \propto ( \sigma_{\rm NXS,30d}^{2} )^{1.353 \pm 0.034}$, with a higher residual variance. This relation allows us to estimate short-timescale variability for sources lacking direct 3-day measurements but with well-constrained monthly light curves.

Using this empirical relation, we extrapolated $\sigma_{\rm NXS, 3d}^{2}$ values for 320 additional sources in our original monthly-variable sample, expanding our 3-day timescale coverage from 87 to 407 AGNs. These extrapolated values were incorporated as variability amplitude scalings in the AGN flux simulations developed for CTAO forecasts and will be detailed in the forthcoming AGN Population paper.


The stability of the $\sigma_{\rm NXS, 3d}^{2}$ vs. $\sigma_{\rm NXS, 30d}^{2}$ correlation was tested under different filtering conditions, including stricter or more relaxed thresholds for Test Statistic (TS), percentage of allowed data gaps, and treatment of unconstrained points. In all cases, the fit behavior remained robust. Preliminary tests suggest the scaling also holds at intermediate cadences, such as weekly binnings, and appears consistent across SED classes and source types (BL Lacs vs. FSRQs), though classification-dependent nuances remain under investigation (Passos Reis et al, in prep.).

This variability scaling provides a valuable tool for modeling AGN behavior at timescales below direct observational limits. By linking long and short-timescale variability, we enable the generation of realistic light curves that better represent the dynamical nature of AGN emission, which is critical for assessing detection probabilities in future CTAO time-domain surveys.

\section{Discussion and Results}

A key aspect of AGN detectability in the VHE regime is their intrinsic variability. Flux variations, especially on short timescales, can significantly increase the chances of detection during a flaring state. In this work, we employed the normalized excess variance ($\sigma_{\rm NXS}^{2}$) to quantify variability across three timescales: monthly, weekly, and 3-day bins. By establishing an empirical scaling relation between the monthly and 3-day $\sigma_{\rm NXS}^{2}$ values, we have extended the characterization of long-term variability to sources with incomplete short-cadence light curves, typically limited by upper limits or unconstrained flux measurements.

To forecast which AGNs CTAO might detect, we relied on the well-characterized GeV sky from Fermi-LAT and extrapolated it into the VHE regime. This effort, part of the CTAO AGN Population Task Force, further discussed in a CTAO Consortium publication, combines spectral energy distributions and variability properties to refine population models. The scaling relation we derived enabled the inclusion of short-term variability amplitudes in sources lacking full 3-day coverage, thus expanding the sample of simulated light curves. Compared to static-flux predictions, incorporating short-timescale variability increases the expected number of detectable AGNs in both 5-hour and 20-hour CTAO exposures.
These findings emphasize the importance of variability-informed modeling in future extragalactic VHE surveys and illustrate how statistical trends across the AGN population can help us to optimize time-domain observation strategies.

Beyond detectability forecasts, our preliminary results offer insights into the physical processes driving AGN variability. High-energy emission is expected to originate in compact regions of relativistic jets, where mechanisms such as stochastic acceleration, shock propagation, or magnetic reconnection may dominate. Variability studies across multiple cadences and AGN SED classes can help constrain these mechanisms by revealing the amplitude and frequency of flaring episodes. The analysis presented here lays the groundwork for a deeper exploration of AGN variability across both timescales and energy regimes, supporting CTAO’s mission to probe the most extreme astrophysical environments.

\end{document}